\documentclass[%
 reprint,
showpacs,
 amsmath,amssymb,
 aps,
 pre
]{revtex4-1}

\usepackage{graphicx}
\usepackage{dcolumn}
\usepackage{bm}
\usepackage{siunitx}
\usepackage{xcolor} 
\usepackage[normalem]{ulem} 

\newcommand{\md}{\mathrm d}
\newcommand{\kB}{k_{\rm B}}
\newcommand{\kT}{\kB T}
\def\F1{${\rm F}_1$}
\def\Fo{${\rm F}_{\rm o}$}
\newcommand{\fix}{}
\newcommand{\fixx}{}
\newcommand{\fixxx}{}
\newcommand{\stkout}[1]{}
\newcommand{\stkoutt}[1]{}
\newcommand{\stkouttt}[1]{}

\begin{document}

\preprint{APS/123-QED}

\title{Modeling work-speed-accuracy trade-offs in a stochastic rotary machine}

\author{Alexandra K.\ S.\ Kasper}
\altaffiliation{Current address: Beedie School of Business, Simon Fraser University, Burnaby, British Columbia, V5A1S6 Canada}
\author{David A.\ Sivak}%
\email{dsivak@sfu.ca}
\affiliation{%
 Department of Physics, Simon Fraser University, Burnaby, British Columbia, V5A1S6 Canada}%

\date{\today}

\begin{abstract}
Molecular machines are stochastic systems that catalyze the energetic processes keeping living cells alive and structured. 
Inspired by the examples of \F1-ATP synthase and the bacterial flagellum, we present a minimal model of {\fixx an externally} driven stochastic rotary machine. 
We explore the trade-offs of work, driving speed, and driving accuracy when changing driving strength, speed, and the underlying system dynamics. 
We find an upper bound on accuracy and work for a particular speed.
Our results favor slow driving when tasked with minimizing the work-accuracy ratio and maximizing the rate of successful cycles.
Finally, in the parameter regime mapping to the dynamics of \F1-ATP synthase, we find a significant decay of driving accuracy at physiological rotation rates, raising questions about how ATP synthase achieves reasonable or even remarkable efficiency \textit{in vivo}.
\end{abstract}

\pacs{05.70.Ln, 05.40.-a, 05.10.Gg, 02.60.Cb}
\keywords{Stochastic physics, molecular machines, Fokker-Planck}
\maketitle

\section{\label{sec:level1}Introduction}
Molecular motors are stochastic machines built from protein-based components and are capable of converting between different forms of energy in the cell. The molecular details of energy transduction in these soft-matter objects are often complex, and relatively few systems have a well-characterized mechanism.  The \F1 subunit of \Fo\F1-ATP synthase and bacterial flagellar motors are the most prominent examples of stochastic rotary molecular machines that have been extensively studied and for which there is reasonable agreement about their structure and function~\cite{Yoshida2001a, Morimoto2014}. 

Both \F1 and flagellar motors are capable of reversible operation, as confirmed by direct observation~\cite{Toyabe2011, Yoshida2001a,Saita2015,Morimoto2014,Delalez2010}.
Flagellar motors switch between clockwise and counter-clockwise rotation in response to changing extracellular conditions in order to control the movement of the bacterial cell, but the ion-transfer process driving the rotation does not switch directions~\cite{Morimoto2014,Delalez2010}. In other words, the same chemical process can drive mechanical rotation in either the clockwise or counter-clockwise direction.

\F1, on the other hand, exhibits reversible mechanochemical coupling.
{\fix \emph{In vivo}, it acts as part of the \Fo\F1 ATP synthase, where the integral membrane complex \Fo\ transports hydrogen ions across a membrane (down their gradient), forcing rotation of a crankshaft, which in turn drives each of the three identical subunits of \F1 to synthesize ATP from ADP and inorganic phosphate (P\textsubscript{i}) against a chemical-potential difference~\cite{Junge:2015fo}.
Single-molecule studies typically remove \Fo, attach a magnetic bead or pair of beads to the crankshaft attached to \F1, and force and/or measure crankshaft rotation using a magnetic trap~\cite{Rondelez:2005be} or electrorotation~\cite{Toyabe2011}, thereby driving} the energetically costly reaction of ATP synthesis~\cite{Yasuda1998,kinosita00,Rondelez:2005be,Toyabe2011}.
Under suitable conditions {\fix (an excess of ATP), \F1}
can also hydrolyze ATP and rotate 
{\fix the crankshaft}
in the opposite direction~\cite{Yoshida2001a,Rondelez:2005be,Toyabe2011}. 
Unlike flagellar motors, the directionality of mechanical rotation and the chemical reactions in \F1 are coupled: when one reverses, so does the other.

Bacterial flagellar motors and \F1 are believed to be highly efficient machines~\cite{Meister1987,Yasuda1998}. 
\F1 is reported to have an efficiency of nearly 100\% and exhibit near-perfect mechanochemical coupling between the crankshaft and chemical catalysis~\cite{Yasuda1998,Toyabe2011,Saita2015,Soga2017}. 
The energy transfer is mediated by the rotation of the central crankshaft via a mechanical torque (caused by the rotation of the \Fo\, subunit or single-molecule manipulation), ultimately producing conformational changes in the {\fix three} catalytic sites of \F1.

Mechanochemical coupling lies in the interdependence of the energetics of the mechanical and chemical states: the energies of mechanical states depend on the chemical state, and vice versa. Mechanochemical coupling can be quantified by the match between the number of rotations and the number of ATP produced~\cite{Toyabe2011}. 
Since \F1 has three catalytic sites {\fixx (spatially arranged with a threefold rotational symmetry)}, perfect mechanochemical coupling can be quantified by three ATP produced per full rotation of the F$_1$ subunit of ATP synthase. 
ATP synthase is believed to operate at hundreds of rotations per second in living cells~\cite{Ueno2005}, yet the experiments demonstrating high efficiency use rotation rates two orders of magnitude slower than this. 

It remains unclear whether ATP synthase and bacterial flagellar motors are capable of high efficiency at physiologically relevant rotation rates. Recent measurements of efficiency throughout the cycle of \F1-ATPase without significant opposing force found a maximum efficiency of 72\%~\cite{Martin2018}. Characterizing the breakdown of efficient energy conversion could inform the design of artificial molecular machines, where energetic efficiency would prove useful in applications ranging across computation and information manipulation~\cite{Zulkowski:2014:PhysRevE}, artificial photosynthesis~\cite{McConnell2014}, and drug delivery~\cite{Peng2017}.

Here we computationally explore the behavior of a simple driven stochastic rotary machine. We present a minimal model inspired by the \F1 subunit of ATP synthase, {\fixx \sout{operating} dynamically driven by external single-molecule manipulation to operate} in the ATP synthesis mode. We find an analytic speed-dependent upper bound in the work-accuracy space that is satisfied across the explored parameter space. In the parameter regime representing \F1, our simulations agree with experiments finding near-perfect operation at rotation rates below 10 Hz; yet, our simulations predict a steep decrease in machine accuracy beyond 10 Hz, with significantly reduced operation by 100 Hz. We believe this result is consistent with experimental results at speeds $\sim$100 Hz when unsuccessful cycles are considered. 

{\fix Previous research in our group~\cite{Lucero:2019gd} studied the effect on work and flux in a simpler discrete-state model of protocols designed using a near-equilibrium theory~\cite{SivakOptimalPath} to vary their angular velocity so as to reduce work.  Here, by contrast, we study in a more accurate continuous-state model the analytic bounds for work-speed-accuracy trade-offs when driving with constant angular velocity, and draw connections with the experimental single-molecule literature on \F1.}

\section{Model}
\label{sec:model}
We propose a minimal thermodynamic model of {\fixx an externally} driven rotary machine, where we tune driving strength, driving speed, and intrinsic machine dynamics to explore the trade-offs between energetic efficiency, speed, and mechanochemical coupling {\fixx (accuracy of driving)}. 

The \F1 system has two prominent coarse-grained state variables: the crankshaft angle and the chemical coordinate quantifying progress of ATP synthesis. Other models of \F1 involve higher-dimensional state variables, for example defining the chemical state as the occupancy of each of the three catalytic states~\cite{Wang1998,Atp2005}. 
In the spirit of minimalism, our model has a single dependent state variable, the chemical coordinate $\theta$, and one independent state variable, the crankshaft angle $\theta_0$. We assume perfect coupling between the crankshaft and the experimental apparatus driving the system {\fixx (which can thus be jointly modeled by the single mechanical coordinate $\theta_0$). We also assume perfect coupling between the distinct chemical states at \F1's three catalytic sites (which can thus be jointly modeled by the single chemical coordinate $\theta$).} 

\begin{figure}
\includegraphics[width=\columnwidth]{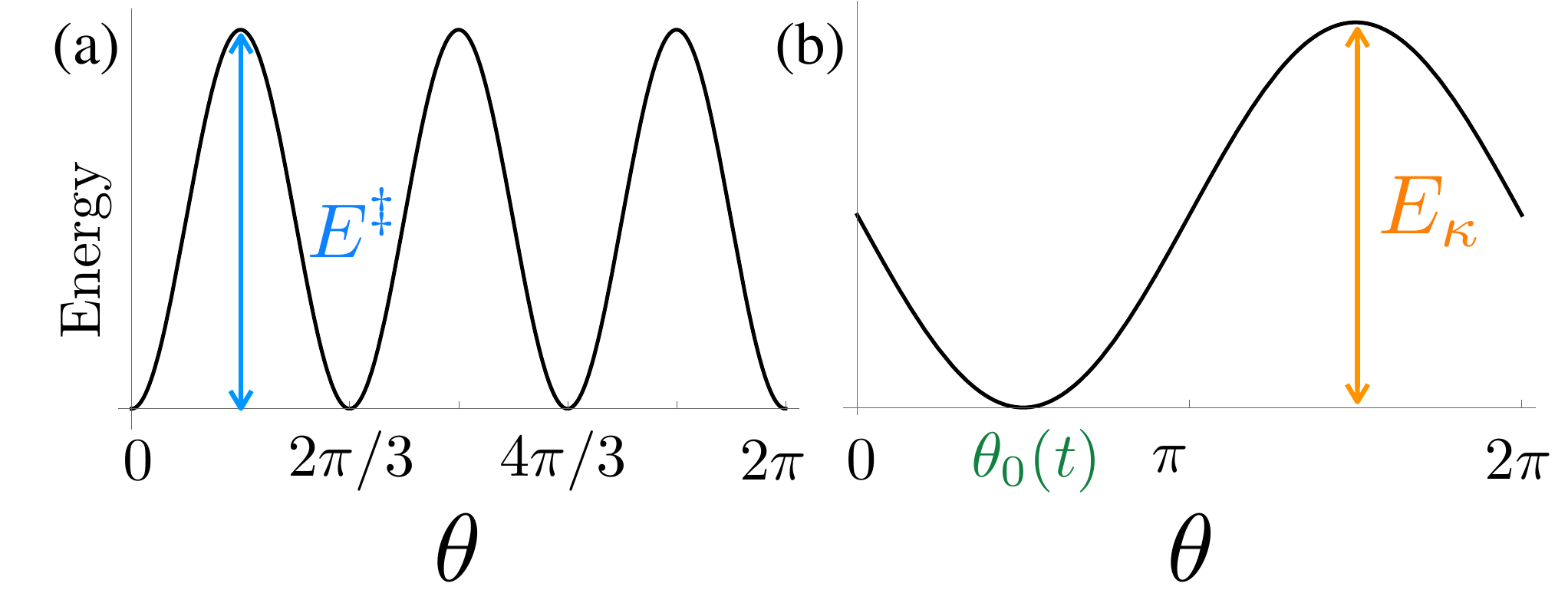}
\caption{The two components of the potential. (a) Molecular potential with three minima and energy barriers of height $E_\ddagger$. (b) Time-dependent driving potential with a single minimum at $\theta_0(t)$ and height $E_\kappa$.}
\label{fig:Potentials}
\end{figure}

This model machine (a continuous analog of the model in \cite{Lucero:2019gd}) has potential energy landscape $U(\theta,t)$ with two sinusoidal components: a molecular potential {\fix (Fig.~\ref{fig:Potentials}(a))},
\begin{equation}
\label{eq:MolecularPotential}
{\fix
U_{\rm molec}(\theta) = \frac{1}{2}E^\ddagger\left(1-\cos 3\theta\right) \ ,
}
\end{equation}
and a time-dependent driving potential {\fix (Fig.~\ref{fig:Potentials}(b))}, 
\begin{equation}
\label{eq:DrivingPotential}
{\fix
U_{\kappa}(\theta,t) = \frac{1}{2}E_\kappa\left\{1-\cos\left[\theta-\theta_0(t)\right]\right\} \ .
}
\end{equation}
The molecular potential {\fix $U_{\rm molec}(t)$} represents the internal energetics of \F1: the three minima at $\theta=0$, $2\pi/3$, and $4\pi/3$ represent the three metastable states. The minima are separated by energy barriers of height $E^\ddagger$, {\fix representing \F1's chemical transition states}. 
This model easily generalizes to a rotary machine with an arbitrary number $N$ of distinct minima and hence distinct steps.

The driving dynamics are modeled by a time-dependent driving potential {\fix $U_\kappa(\theta, t)$} with a single minimum at the crankshaft angle $\theta_0(t)$, the \emph{control parameter} in our model. $\Theta$ denotes the protocol $\theta_0(t)$ (the values $\theta_0$ takes {\fix as a function of time}), consisting of a complete cycle driving through one full rotation over the period $\tau$. {\fix The driving speed is tuned by changing the cycle period.}
$E_\kappa$ is the peak height of the driving potential and thus quantifies the strength of driving. {\fix The driving potential represents the time-dependent rotation of a magnetic trap at constant velocity, commonly used for the external manipulation of \F1 \textit{in vitro}~\cite{Rondelez:2005be}.}

\section{Methods}
\label{sec:methods}
We used the Smoluchowski equation (the overdamped Fokker-Planck equation) to simulate system dynamics subject to driving. The Smoluchowski equation for the evolution of the probability distribution over the chemical coordinate $\theta$ is~\cite{Risken1989} 
\begin{equation}
	\label{eq:System}
	\frac{\partial P(\theta,t)}{\partial t}=\frac{D}{\kT}\frac{\partial[U'(\theta,t) P(\theta,t)]}{\partial \theta}+D\frac{\partial^2P(\theta,t)}{\partial \theta^2} \ ,
\end{equation}
for angular diffusion coefficient $D$, temperature $T$, and Boltzmann's constant $\kB$. 

The average system energy is
\begin{equation}
\label{eq:Energy}
E(t)=\int_0^{2\pi}\md\theta \, P(\theta,t) U(\theta,t) \ .
\end{equation}    
The work is defined as the energy change when the potential is updated, and the heat as the energy change when the system relaxes on a static landscape. Because of the discrete nature of simulations, these two processes are distinguishable substeps within one full time update. The system returns to the same control state at the end of the driving cycle, so the equilibrium free energy change is zero; thus, all work over one cycle is necessarily excess work, above and beyond the minimum energy input required to change the equilibrium state of the system.

The flux $J(\theta,t)$ is the net instantaneous probability flow at a particular angle and time, 
\begin{equation}
\label{eq:flux}
J(\theta,t)=-D\frac{U'(\theta,t)}{\kT}P(\theta,t)-D\frac{\partial}{\partial \theta}P(\theta,t).
\end{equation}
We define accuracy $\eta$ as the net probability flow over one cycle, averaged over angle:     
\begin{equation}
\label{eq:accuracy}
\eta=\frac{1}{2\pi}  \int_0^{2\pi} \int_{0}^{\tau}J(\theta,t) \, \md t \, \md \theta \ .
\end{equation}

Accuracy quantifies the mechanochemical coupling, the probability that the rotating crankshaft successfully drives a full complement of three chemical reactions per driving cycle {\fixx (maximal chemical output of \F1)}. {\fixxx Accuracy represents a mechanochemical efficiency, the efficiency with which directional mechanical motion of the crankshaft is converted into directional progress of the chemical degree of freedom.} {\fixx Accuracy can also be thought of as the absence of ``slippage'', as conceived for linear motors in \cite{ChowdhuryReview} -- a decrease in accuracy corresponds to an increase in slippage between the crankshaft angle and the chemical coordinate, i.e. the chemistry is not tightly coupled to the mechanics.

The accuracy is also the chemical output per cycle, and thus} accuracy divided by the cycle time gives the average {\fixx(chemical)} flux over one cycle: 
\begin{equation}
\label{eq:avgFlux}
\langle J \rangle_{\Theta} = \frac{\eta}{\tau} \ .
\end{equation}

We evolved Eq.~\eqref{eq:System} using finite-difference numerical methods~\cite{Recipes}. To exclude transient behavior resulting from system initialization, the probability distribution was evolved for many driving cycles, continuing until reaching periodic steady state (PSS), which is defined by a complete driving cycle leaving unchanged the system's angular distribution: $P(\theta,t)=P(\theta,t+\tau)$ for all angles $\theta$ and times $t$ over a cycle duration $\tau$. 
This implies no net energy accumulation in the system: all energy entering as excess work is dissipated as heat over a complete cycle.

We defined the time scale for all simulations using an estimation of the diffusion coefficient of \F1 following \cite{Xu2008}. 
We determined, based on observed angular occupancy probabilities in \cite{Toyabe2011}, that our model best maps to \F1's energetics when $E^\ddagger=2\, {\fix \kT}$. 
See App.~\ref{sec:tauDetermination} and \ref{sec:eDEstimate} for calculation details.

\section{Results}
\begin{figure}[htbp]
\includegraphics[width=\columnwidth]{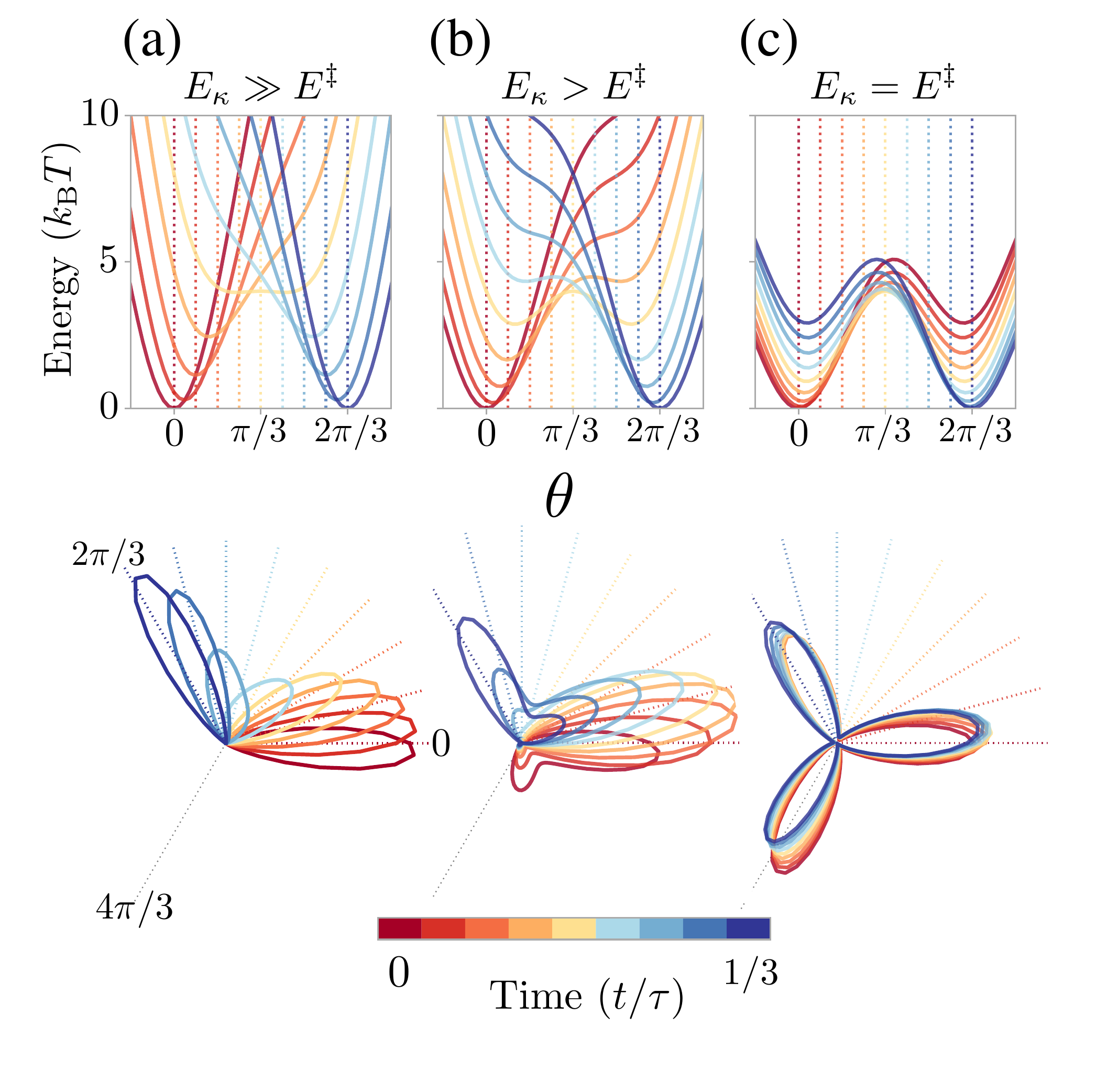}
\caption{\fix Instantaneous energy potentials (top) and the corresponding PSS angular distributions $P(\theta,t)$ (bottom),
sweeping over a third of a driving cycle ($\theta_0$ changing from 0 to $2\pi/3$ radians) as the color changes from red to blue. Colored dashed straight lines show the corresponding instantaneous value of $\theta_0$. 
The rotation rate is 14 Hz and barrier height $E^\ddagger=4 \ \kT$. 
The driving strength $E_{\kappa}$ is (a) $32 \ \kT$, (b) $16 \ \kT$, or (c) $4 \ \kT$. The PSS angular distributions are shown on polar plots with the radius corresponding to the probability, and normalized within a given plot but not between plots.} 
\label{fig:ProbDists}
\end{figure}

Figure~\ref{fig:ProbDists} shows one third of a PSS cycle. 
Under strong driving ($E_\kappa \gg E^\ddagger$; Fig.~\ref{fig:ProbDists}a), $P(\theta,t)$ has a single peak closely aligned with $\theta_0$. The decreased peak height for $\theta_0 \sim \pi/3$ represents the spread of the distribution as the system awaits a large thermal fluctuation to kick it over the barrier of the molecular potential. 
At intermediate driving ($E_\kappa > E^\ddagger$; Fig.~\ref{fig:ProbDists}b), each distribution has two main peaks, indicating significant probability in two minima of the molecular potential. These visualizations for large and intermediate driving exhibit the qualitative behavior of a cycle with high accuracy: most of the probability flows through the system over the course of one driving cycle. For weak driving ($E^\ddagger\gtrsim E_\kappa$; Fig.~\ref{fig:ProbDists}c), comparable probability remains in each minimum due to the rarity of thermal kicks with sufficient strength to carry the system over the large barrier. 
This shows low accuracy: driving has a weak effect on the distribution. 

\begin{figure*}[htbp!]
\includegraphics[width=0.8\textwidth]{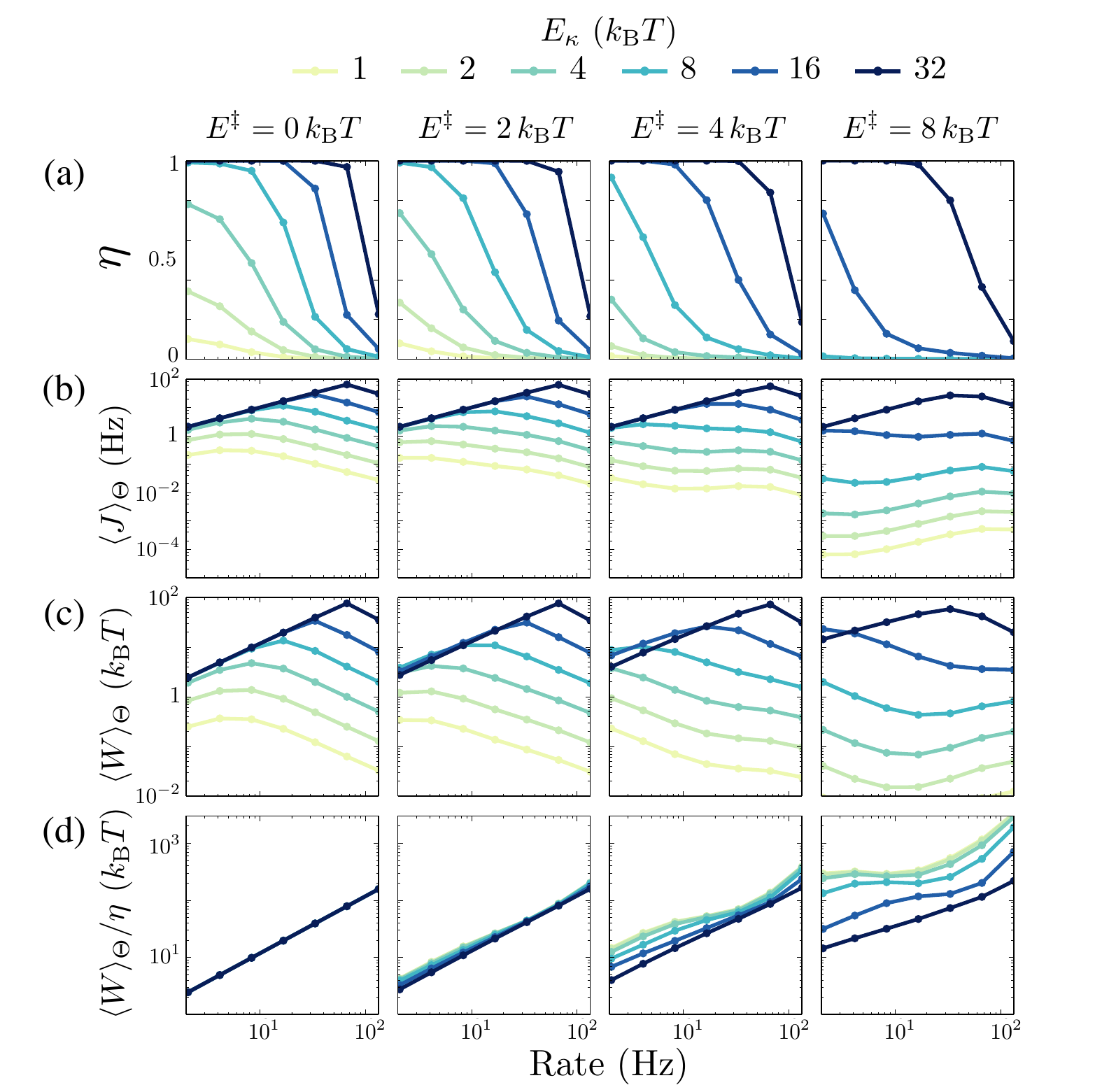}
\caption{(a) Accuracy, (b) average flux, (c) work per cycle, and (d) work/accuracy ratio as a function of rotation rate.
(a) Accuracy decreases as cycle rate increases. (b) Flux is maximized at intermediate rotation rate. (c) Excess work per cycle peaks at intermediate rotation rate. (d) The ratio of excess work and accuracy increases with both rotation rate and barrier height $E^\ddagger$.
Molecular barrier height $E^\ddagger$ varies between sub-plots and driving strength $E_\kappa$ varies within each sub-plot.}
\label{fig:NaiveLines}
\end{figure*}

Figure~\ref{fig:NaiveLines} explores the effect of driving rate on accuracy, cycle-averaged flux and work, and the ratio of work and accuracy.
Figure~\ref{fig:NaiveLines}a shows that accuracy decreases monotonically with rotation rate for all examined $E^\ddagger$ and $E_\kappa$: a rapidly rotating crankshaft reduces the probability that the chemical coordinate traverses a cycle before the crankshaft returns to its original angle.

Since a fast machine with a low accuracy may still produce more product per unit time than a slower but more accurate machine, we also examine the flux per cycle. Figure~\ref{fig:NaiveLines}b shows that for some combinations of $E^\ddagger$ and $E_\kappa$, flux per cycle peaks at intermediate rotation rate. These peaks coincide with the region where $\eta$ sharply decreases with increasing rotation rate (Fig.~\ref{fig:NaiveLines}a), indicating that maximal machine output occurs at intermediate accuracy and rotation rate. 

Work per cycle also peaks at intermediate rotation rate when $E_\kappa>E^\ddagger$ (Fig.~\ref{fig:NaiveLines}c). The peak work occurs at the same rotation rate as the peak flux, meaning maximal output maximizes the cost per cycle. For a fixed energy budget per cycle, the peak in work indicates that there will often be two cases that satisfy this constraint: a faster yet less accurate machine, or a slower yet more accurate machine. 

Figure~\ref{fig:NaiveLines}d shows the work per unit probability driven through one cycle, $\langle W \rangle_\Theta/\eta$. 
Intuitively, $\langle W \rangle_\Theta/\eta$ is the cost of a successful cycle, obtained by normalizing the excess work by the total probability flow. When $E^\ddagger=0$, this cost increases linearly with rotation rate and independently of $E_\kappa$, consistent with analytic predictions~\cite{Mazonka1999}. 

\begin{figure*}[htbp]
\includegraphics[width=0.8\textwidth]{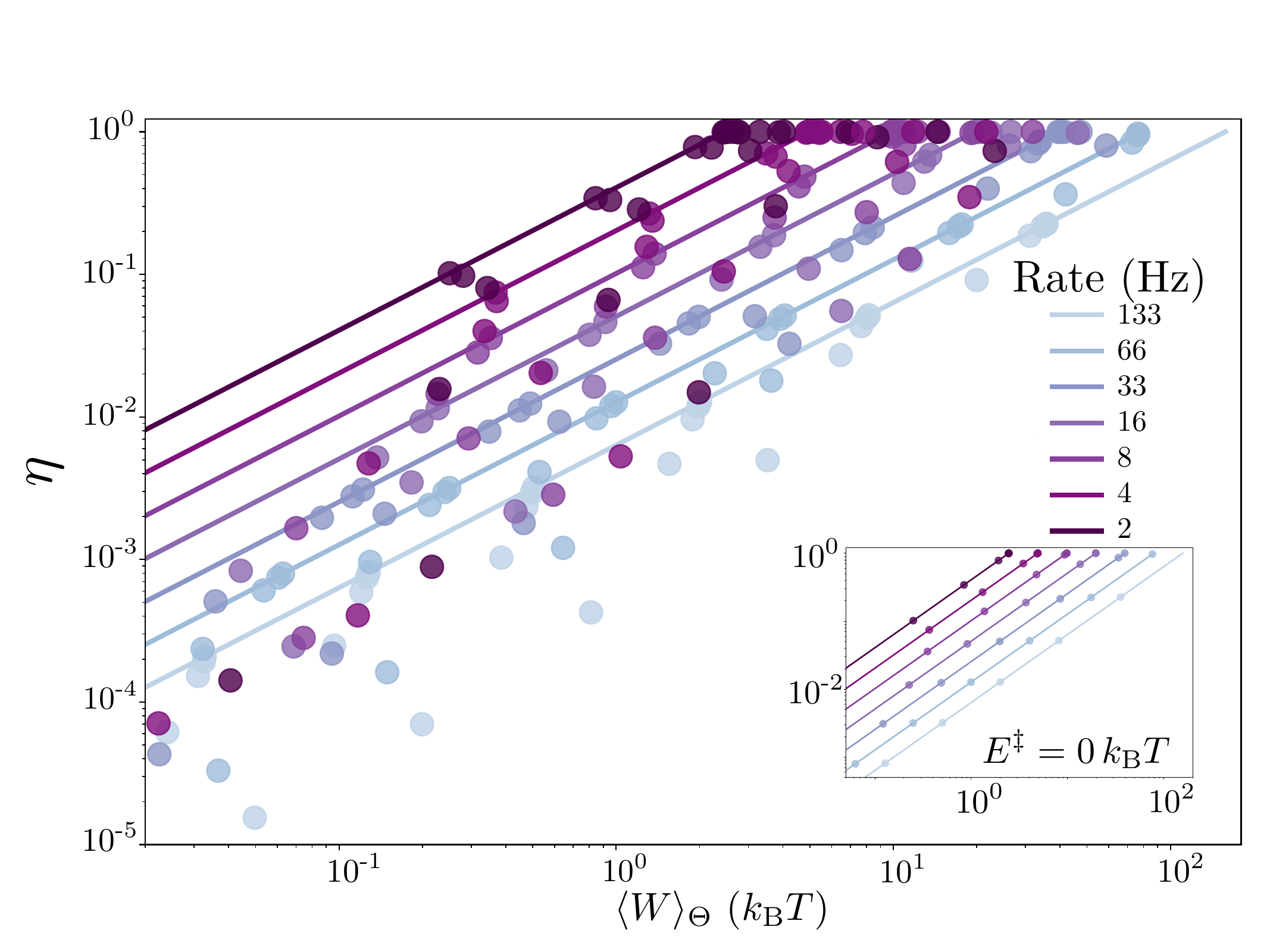}
\caption{Accuracy and work per cycle for varying rotation rate $v=2\pi/\tau$, barrier height $E^\ddagger$, and driving strength $E_\kappa$. Color indicates rotation rate, while points of the same color span all combinations of parameter values $E^\ddagger$ = 1,2,4,8 $\, {\fix \kT}$ and $E_\kappa$ = 1,2,4,8,16,32 $\, {\fix \kT}$. Solid lines are independently calculated from theory~\eqref{TheoryLine}, lying above all corresponding numerical calculations, describing a performance frontier for a particular rotation rate. Decreasing the rotation rate pushes the frontier towards higher accuracy and lower work per cycle. Inset: numerical calculations for flat molecular potential ($E^\ddagger=0$) coincide with theory lines~\eqref{TheoryLine}.}
\label{fig:FlowDots}
\end{figure*}

Figure~\ref{fig:FlowDots} shows a parametric plot of accuracy $\eta$ versus $\langle W \rangle_\Theta$, permitting examination of the three-way trade-off between rotation rate, accuracy $\eta$, and work per cycle $\langle W \rangle_\Theta$. For each color, rotation rate is held fixed while $E_\kappa$ and $E^\ddagger$ are varied. Generally speaking, machine function improves up (increasing accuracy) and to the left (decreasing work per cycle). 

We find a boundary of operation that is independently calculated from the system's physical parameters, with no fitting parameters. 
For a particular driving rate, all explored $E^\ddagger$ and $E_\kappa$ produce accuracy lying below these lines. The work per cycle for the upper terminus of each line
is calculated from 
\begin{subequations}
\begin{align}
\langle W \rangle_{\Theta}^* &= v^2 \frac{\kT}{D}t \\
&= \frac{(2\pi)^2}{\tau}\frac{\kT}{D} \ ,
\label{TheoryLine}
\end{align}
\end{subequations}
the work imposed on a diffusing particle by a quadratic trap translating at constant velocity $v = 2\pi/\tau$ over duration $\tau$~\cite{Mazonka1999}\footnote{Here we have used the Einstein relation to substitute $\gamma$ from the original paper with $\frac{\kT}{D}$.}.

To obtain the boundary line for each rate, we assume that work decreases linearly with slippage, i.e., half the accuracy requires half the work. Each line describes a decrease in work that is proportional to the decrease in accuracy: $\langle W \rangle_{\Theta} = \eta \langle W \rangle_{\Theta}^*$. $\eta$ has an upper bound of unity. This assumption only strictly holds for a flat molecular potential with $E^\ddagger=0$: across varying $E_\kappa$ and rotation rate, any simulation with no molecular energy barriers ($E^\ddagger=0$) falls exactly on the respective theory line (Fig.~\ref{fig:FlowDots} inset).

\section{Discussion}
We explored the fundamental limits of a driven stochastic rotary machine by developing a minimal model of driving inspired by the \F1 subunit of the molecular machine ATP synthase. 
Chemical output (accuracy) per cycle is increased by slower driving, and there is reason to believe the system cannot maintain perfect mechanochemical coupling at arbitrarily high speeds. 
For a particular energy (work) budget, there are often two accuracy-speed pairs that satisfy the constraint: a faster but less accurate option, and a slower but more accurate option -- both with the same flux (chemical output per unit time).

The theory line in Fig.~\ref{fig:FlowDots} is a Pareto frontier of operation~\cite{Shoval2012}.
We expect that no combination of $E^\ddagger$ or $E_\kappa$ (molecular or driving energetics, respectively) can both increase accuracy and decrease work. This is because the work cost $\langle W \rangle_\Theta/\eta$ per successful cycle increases with $E^\ddagger$, as seen in Fig.~\ref{fig:NaiveLines}d: it costs more to successfully drive the machine through its cycle when there are molecular barriers, compared to a flat molecular landscape.

An important feature of our simulations is that there is no net accumulation of energy in the system upon completing one PSS cycle: the net excess work is equal to the amount of heat released -- \emph{over one complete cycle}. However, the \emph{instantaneous} flow of work and heat are not balanced in general. 
Rather, only in the quasistatic driving case is driving sufficiently slow that the system remains in equilibrium: work is dissipated as heat instantaneously. But given the finite driving rates of our simulations, the system is out of equilibrium. In fact, it is often far from equilibrium, and driving can (at least transiently) result in negative work: the update to the potential decreases the energy of the system. 

The negative work implies the system is able to retain excess work and return some of it later rather than releasing it all as heat. 
When the molecular potential is not flat ($E^\ddagger > 0$), we hypothesize that more heat is released during relaxation on average, resulting in less stored energy that can be recovered later as negative work and therefore a higher net work cost. 

For large $E_\kappa$, the driving potential dwarfs the molecular potential, resulting in a system that essentially does not feel its molecular potential and falls close to the frontier.

For the \F1-like system ($E^\ddagger=2\, {\fix \kT}$), we only found near-perfect accuracy (an indicator of tight mechanochemical coupling) for the slowest driving. Single-molecule experiments reporting near-perfect efficiency of ATP synthase have rotational driving rates on the order of 0.1-10 Hz~\cite{Toyabe2011,Yasuda1998, Soga2017, Saita2015}. 
In our model, we find near-perfect performance at 2 and 4 Hz for driving strengths of $E_\kappa = 16$ and $32\, {\fix \kT}$.
ATP synthase is believed to operate at hundreds of Hz in living cells~\cite{Ueno2005}, and \F1 is capable of rotating at 700 Hz with gold beads attached to the $\gamma$ subunit~\cite{Nakanishi2006}. In our system, the accuracy is greatly diminished above 10 Hz, with system response (accuracy) falling to near zero for 100 Hz driving at all driving strengths explored.

{\fixxx If one were to relax our assumption of tight coupling between the external manipulation and crankshaft rotation, then our calculated accuracy would represent an upper bound (and the speed at which our calculations show a drop in mechanochemical efficiency would represent an upper bound) on that for a more loosely coupled system.}

Experimentalists have noted that in higher-speed regimes the rotation rate as predicted by measuring ATP hydrolysis or synthesis rates from bulk solution is lower than the individual rotation rates observed by tracking beads attached to the crankshaft~\cite{Nakanishi2006}. Furthermore, the average rotation rate was more than 10$\times$ faster than that predicted from the hydrolysis rate and an assumed 3 ATP molecules hydrolyzed per rotation~\cite{Nakanishi2006}. \cite{Yasuda2001} found a similar mismatch.
\cite{Nakanishi2006} attributes the discrepancy---at high rotation rate---between rates of rotation and hydrolysis to temporary inactivation of up to 90\% of the \F1 at any given time. However, their calculation of the average rotation rate factored in the paused beads and still produced a rate nearly 13$\times$ faster than expected from the chemical outputs in the bulk. 
Our results for $E^\ddagger=2\, {\fix \kT}$ find an accuracy $<$20\% over 100 Hz, which can be interpreted as over 80\% of driving cycles being unsuccessful at high speeds. 

Future extensions of this work could add features to the model (\S\ref{sec:model}) to more closely capture the \emph{in vivo} physical operation of ATP synthase. 
{\fix We modeled experimental conditions involving equilibrium concentrations of ATP, ADP, and P\textsubscript{i}, where there is no chemical potential difference associated with ATP synthesis or hydrolysis.  Perhaps the simplest way to model the effect of nonequilibrium chemical concentrations (and hence nonzero chemical potential difference over a cycle) would be to introduce a constant force, equivalent to a linear tilt of the energy landscape. The product of this resisting force and the flux would quantify the power output, permitting calculation of a thermodynamic efficiency.}
{\fix One could also replace} 
the magnetic trap with a fluctuating rotary component (modeling \Fo) viscoelastically coupled to \F1 and driven by a force due to cross-membrane proton concentration differences {\fix (establishing constant-affinity kinetics and permitting application of the thermodynamic uncertainty relation~\cite{barato15,barato15b})}. 
{\fix These changes}
would give a useful model system to explore the physical constraints on energy flows between such strongly coupled nonequilibrium subsystems~\cite{Crooks:2019bk}, and the resulting design principles~\cite{Large:2018dh} governing the architecture of such machines. 
Beyond the potential implications for \textit{in vivo} \F1 and other naturally evolved machines, consideration of our entire parameter space gives a broad picture of this class of stochastic rotary machines, potentially informing the design of artificial molecular motors.  

\begin{acknowledgments}
The authors thank John Bechhoefer, Nancy Forde, and Joseph Lucero (SFU Physics) for illuminating conversations and insightful comments on the manuscript. This work was supported by a Natural Sciences and Engineering Research Council of Canada (NSERC) CGS Masters fellowship (A.K.S.K.), a C.\ D.\ Nelson Multi-Year Funding fellowship (A.K.S.K.), an NSERC Discovery Grant (D.A.S.), a Tier-II Canada Research Chair (D.A.S.), and was enabled in part by support provided by WestGrid (\href{www.westgrid.ca}{www.westgrid.ca}) and Compute Canada Calcul Canada (\href{www.computecanada.ca}{www.computecanada.ca}).
\end{acknowledgments}

\appendix
\section{Equating physical time and simulation time}
\label{sec:tauDetermination}
The time elapsed in the simulation must be mapped to meaningful time units in order to compare with observed behavior of \F1. The diffusion coefficient is used to map between simulation time and real time. The Einstein relation, $D=\kT/m\gamma$, allows for a direct conversion by equating the experimental diffusion coefficient, $D_\textrm{exp}$, with the diffusion coefficient defined by the parameters $\kT$ and $m\gamma$.

In an experiment using a magnetic bead tethered to the crankshaft, the experimental frictional drag coefficient $\gamma_\textrm{exp}$ is
\begin{equation}
m\gamma_\textrm{exp}=8\pi\eta r^3+6\pi\eta y^2 r \ ,
\end{equation}	
where $\eta_{\textrm{H}_2\textrm{O}}=10^{-9} $ pN s nm$^{-2}$ is the viscosity of the surrounding liquid, $r$ is the radius of the bead, and $y$ is the distance between the center of the bead and the rotational axis~\cite{Xu2008}\footnote{The published equation is missing the factor of $r$ in the second term.}.  
Using reasonable values for $r$ and $y$ ($y=r=287\,\si{nm}{\fix /2}$)~\cite{Toyabe2011} and room temperature ($\kT$=4.114 pN nm), $D_\textrm{exp}$ can be estimated as:
\begin{equation}
D_\textrm{exp} = \frac{\kT}{m\gamma} \approx 30 \,\si{rad^2/s} \ .
\end{equation}
The simulation timestep is identified with a physical time by equating the simulated diffusion coefficient with $D_{\rm exp}$.


\vspace{3ex}

\section{Estimate of $E^\ddagger$ for \F1}
\label{sec:eDEstimate}
The experimentally measured angular distribution of the system can be used to estimate the height $E^\ddagger$ of the intrinsic molecular barriers. Toyabe \textit{et al}. report a standard deviation of $\sigma =20^\circ=0.35$ rad around each well, and confirmed agreement with the Einstein relation $k_\textrm{width}=\kT/\sigma^2$, leading to a value of $k_\textrm{width} \approx 8 \, \kT$. 
A quadratic trap with a spring constant of $k_\textrm{width}$ maps in the present model to a value of $E^\ddagger=\frac{2}{9}k_\textrm{width}$. 
Therefore $E^\ddagger \approx 2\,\kT$ maps the present model to the \F1 system.


%

\end{document}